%% file: ms.tex
\documentclass[12pt,preprint]{aastex}

\shorttitle{Hard X-ray emission of the microquasar GX 339-4 in the low/hard state.}
\shortauthors{Joinet et al.}

\begin{document}
\title{Hard X-ray emission of the microquasar GX 339-4 in the low/hard state.}
\author{A. Joinet\altaffilmark{1}, E. Jourdain\altaffilmark{1}, J. Malzac\altaffilmark{1}, J.P. Roques\altaffilmark{1}, S. Corbel\altaffilmark{2}, J. Rodriguez\altaffilmark{2} and E. Kalemci\altaffilmark{3}}
\altaffiltext{1}{CESR/CNRS-UPS, 9 Avenue du Colonel Roche, BP4346, 31028 Toulouse, France}
\altaffiltext{2}{Universit\'e Paris 7 Denis Diderot and Service d'Astrophysique,
UMR AIM, CEA Saclay, F-91191 Gif sur Yvette, France.} 
\altaffiltext{3}{Sabanc\i University, Orhanl\i - Tuzla, \.Istanbul, 34956, Turkey}

\begin{abstract}

We present the analysis of the high energy emission of the Galactic black
hole binary GX 339-4 in a low/hard state at the beginning of its 2004 outburst.
The data from 273 ks of \textit{INTEGRAL} observations, spread over 4 weeks, are analyzed, along
with the existing simultaneous \textit{RXTE} HEXTE and PCA data. 
During this period, the flux increases by a factor of $\simeq$ 3, 
while the spectral shape is quite unchanged, at least up to 150 keV.
The high energy data allows us to detect the presence of a high energy cut-off, 
generally related to thermal mechanisms, and to estimate the plasma parameters
in the framework of the Comptonization models. We found an electron temperature
of  60-70 keV, an optical depth around 2.5, with a rather low reflection factor
(0.2-0.4). In the last observation, we detected a high energy excess above
200 keV with respect to thermal Comptonization while at lower 
energy, the spectrum is practically identical to the previous
one, taken just 2 days before. 
This suggests that the low and high energy components have a different origin.

\end{abstract}

\keywords{stars: individual: GX 339-4 - gamma rays : observations - black hole physics - accretion, 
accretion discs - X-rays : Binaries}

\section{Introduction}
Since its discovery more than 30 years ago by Markert et al. (1973),
the X-ray binary GX 339-4 has been extensively studied with several 
optical, infrared, X and $\gamma$-ray observatories. The optical companion
is undetectable (Shahbaz, Fender and Charles 2001) but upper limits on 
its luminosity allowed the source to be
classified as a low mass X-ray binary
which could belong to the Galactic bulge (with d$_{min}\ga$6.7 kpc, Zdziarski et al. 2004).
Hynes et al. (2003) 
propose to classify GX 339-4 as a black hole with a mass function of 5.8 M$\sun$.\\
The source harbours a complex variability behavior
with a nearly persistent outbursting X and $\gamma$-ray activity and 
the presence of quite long periods
of quiescence (Kong et al. 2002).
In the soft X-rays 
it was observed  at high-resolution with \textit{Chandra} and \textit{XMM} (Miller et al., 2004a, b) while
Belloni et al. (2005) reports on the source timing and color analysis.
At higher energy its emission has been studied by all the major observatories 
as \textit{GRANAT} SIGMA (Grebenev et al. 1993, Bouchet et al. 1993), \textit{CGRO}  (Grabelsky et al. 1993). Zdziarski et al. (2004) present an extensive spectral analysis of the source 
covering the 1987-2004 period using \textit{CGRO} BATSE, \textit{GINGA} ASM and \textit{RXTE} data.
Homan et al. (2005)  present a coordinated multi-wavelength study of the 2002 outburst and suggest
a non-thermal/jet origin for the optical/near-infrared emission (Corbel \& Fender 2002) in the hard state while the accretion
disk dominates in the soft state.

  In transient galactic black holes such as GX 339-4,  the accretion rate varies 
by several orders of
magnitude. They consequently exhibit a complex spectro-temporal variability.
X-ray monitoring campaigns (with e.g. \textit{RXTE}) demonstrated the existence of   
 two main X-ray spectral states.
Sources are usually
observed with a X-ray luminosity lower than a few percents of the Eddington luminosity in the so called
Low Hard State (hereafter LHS), characterised by a
relatively low flux in the soft X-rays ($\la$ 1 keV) and a high flux
in the hard X-rays ($\sim$100 keV). In the LHS, the high-energy
spectrum can be roughly described by a power-law with spectral
index $\gamma$ varying in the range 1.4-2.1, and a nearly exponential
cut-off at a few hundred keV (see e.g. Gierli\'{n}ski et al. 1997).  
When the luminosity exceeds a few percents of Eddington  the sources can switch to the High Soft State (HSS). The high-energy
power-law is then much softer ($\gamma \ge 2.4$), without any hint for a high energy cut-off and the bolometric
luminosity is dominated by the disc thermal component peaking at a
few keV.  Beside the LHS and HSS, there are several other spectral states that often appear, but not always, when the source is about to switch from one of the two main states to the other. Those states are more complex and difficult to define. We refer the reader to McClintock \& Remillard (2006) and Belloni et al. (2005) for two different  spectral classifications based on X-ray temporal as well as spectral criteria and radio emission (Fender 2006). The hard power law plus cut-off spectrum of the LHS is usually interpreted  as thermal comptonisation in a hot (kTe$\sim$ 100 keV)  optically thin plasma (the corona). In addition to the dominant comptonisation  spectrum there are  several other less prominent spectral features:
\begin{itemize}
\item  a weak soft component associated to the thermal emission  of a geometrically thin optically thick  disk is occasionally detected below 1 keV  as observed in Cygnus X-1 (Balucinska-Church et al. 1995) or in XTE J1118+480 (McClintock et al. 2001, Chaty et al. 2003).
\item  a Fe K$\alpha$ line at $\sim$ 6.4 keV and a Compton reflection bump peaking at $\sim$ 30 keV are believed to be the signature of the irradiation of the cold optically thick disc by the hard X-rays from the corona (George \& Fabian 1991). 
\item   a high energy ($\ge$ 200 keV) excess with respect to simple thermal Comptonization models was detected  in GX-339-4 by OSSE (Johnson et al. 1993) during 
the LHS of its 1991 outburst. A similar excess was detected by COMPTEL in the LHS of Cygnus X-1 (McConnell et al. 2000).  This excess is possibly produced by a tiny population of non-thermal electrons that could be, or not, part of the coronal plasma. With \textit{INTEGRAL} SPI, we have the opportunity to monitor the excess above 200 keV which, as discussed in section 4, is crucial to address the emission mechanism at these energies.
\end{itemize}

In this paper, we will focus only on observations performed when GX 339-4 was in the LHS, while
Belloni et al. (2006) study the evolution of the high energy 
cutoff during the transition from the low/hard to the high/soft state.
We will present here the overall available 
 \textit{INTEGRAL} SPI+IBIS and simultaneous \textit{RXTE} PCA+HEXTE data (see Table~\ref{tab:table1}) 
 which cover the rising phase
 of the 2004 outburst.

The analysis is based on 
the data registered from MJD 53057 (\textit{INTEGRAL} revolution 166 on 2004 February 4) 
to MJD 53085 (\textit{INTEGRAL} revolution 175 on 2004 March 21) limiting us to the low/hard state study.\\
We will briefly describe the \textit{INTEGRAL} and \textit{RXTE} telescopes as well as data analysis methods in Section 2. 
  
 The results, which concern mainly spectral evolution, are presented in Section
3, then discussed in Section 4.

\section{Observations and data analysis}

\begin{figure*}%[!t]
\begin{center}
\includegraphics[scale=0.8]{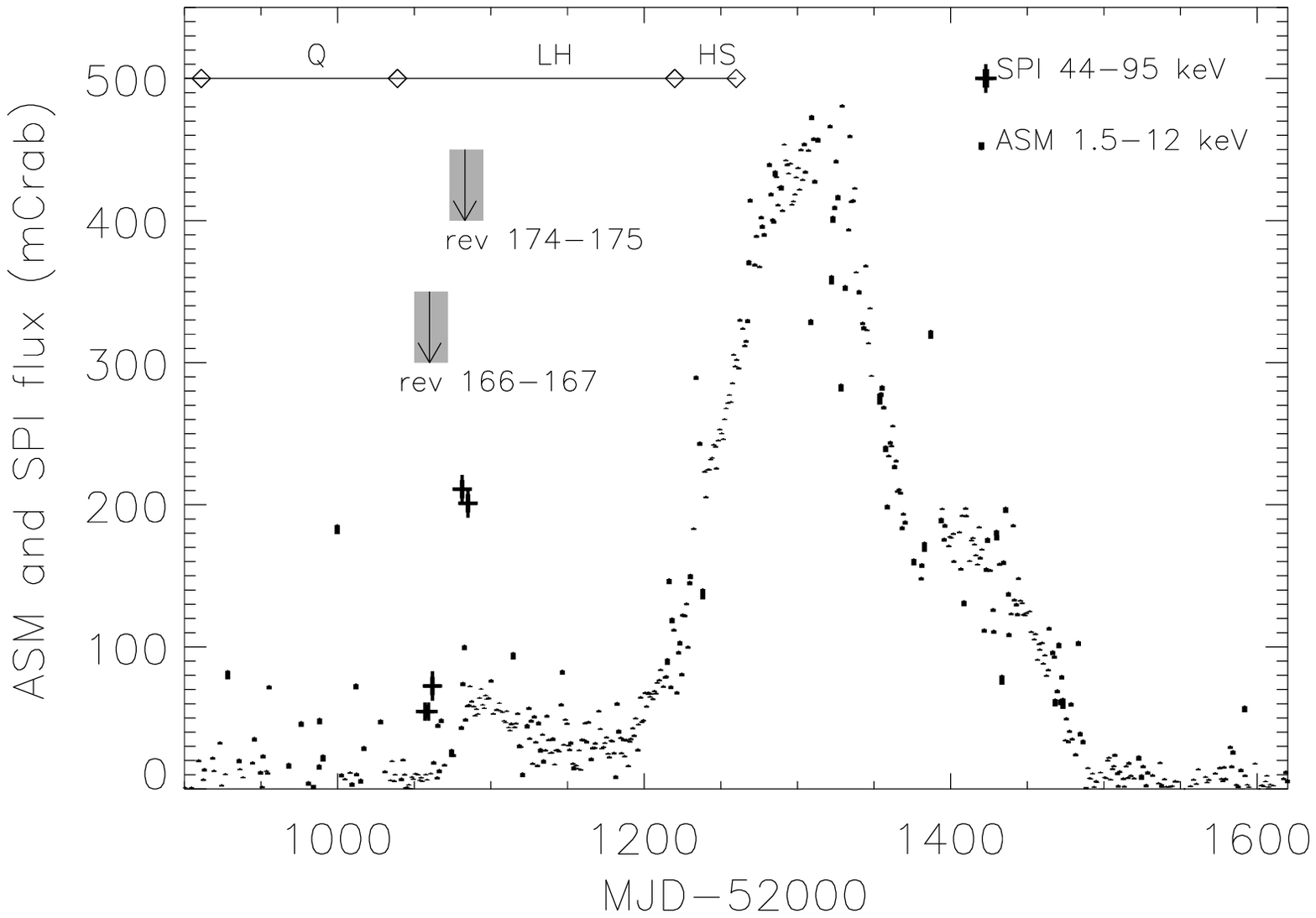}
\end{center}
%\hspace{2cm} 
%\vspace{-0.4cm}
\caption{\textit{RXTE} ASM and SPI light curves of GX 339-4 showing a quiescent period followed
by the 2004 outburst. The different states harboured by the source are 
summarized on the graph : Q=quiescent, LH=low/hard and HS=high/soft (see Remillard 2005).
The arrows represent the \textit{INTEGRAL} observation periods (revolutions
166, 167, 174 and 175).} 
\label{fig:fig3}
\end{figure*}

The first indication of a new reactivation of GX 339-4 has been provided by Buxton et al. (2004) who reported
a radio and optical source fluxes increase on 2004 February 4-5. Then on 2004 February 9,
the X-ray activity has definitively renewed (Smith et al. 2004, Belloni et al. 2004)
and the source was detected by \textit{INTEGRAL} on 2004 February 19 (MJD 53054) (Kuulkers et al. 2004).
This paper is based on data taken during the first part of the outburst when the
source was in the hard state. 
Table~\ref{tab:table1} gives the details of each \textit{INTEGRAL} revolution and corresponding
quasi simultaneous \textit{RXTE} observations used in this analysis (i.e. for the LHS).
Figure~\ref{fig:fig3} shows the \textit{RXTE} ASM light curve of the source in 
the 1.5-12 keV energy range together with \textit{INTEGRAL} observation periods.

\subsection{\textit{INTEGRAL}}
\subsubsection{SPI}
SPI (Spectrometer for \textit{INTEGRAL}, Vedrenne et al. 2003, Roques et al. 2003) is one of \textit{INTEGRAL}'s
two main instruments. Working in the 20 keV - 8 MeV energy 
domain with 19 hexagonal germanium detectors, it possesses an excellent energy 
resolution and a 16$^\circ$ (corner to corner) field of view. 

SPI's imaging
capability is limited with a $2.5^{\circ}$ angular resolution using
 a HURA (Hexagonal Uniform Redundant Array) coded mask, whose cells have 
the same size as the individual detectors. Due to the small number of 
detectors (pixels), SPI image reconstruction methods are based on a combination
of data from several pointings separated by 2$^\circ$ and covering
the same sky region ("dithering strategy", see Jensen et al. 2003).

During the  $\simeq$ 3 day \textit{INTEGRAL} revolution, the observing schedule consists 
of fixed pointings lasting approximatively 30-40 minutes, with a complete 
dithering pattern made up of 25 
pointings (in 5$\times$5 rectangular mode, except during 
the observation program of the GCDE where the pattern is specific) separated by a $2^{\circ}$ angular
distance. 
This method increases the amount of data  in excess of the  number
of unknowns, and allows to better determine the background and 
the position and the flux of sources on the detector plane. 

The signal recorded by the SPI camera on the 19 Ge detectors is composed of 
the contributions from each source in the
 field of view through the instrument aperture, plus the background, which 
 comes mainly from the interaction of high energy particles (
 from cosmic rays or due to solar activity) with the instrument.\\
For $N_s$ sources present in the field of view,  
the data $D_{p}$ obtained during a pointing $p$
for a given energy band can be expressed by the relation:

\begin{equation}
  D_{p} = \sum_{i=1}^{N_s} {R_{p,i} \circledast S_{p,i} } + B_{p} 
\end{equation}

where $R_{p,i}$ is the response of the instrument for the source $i$,
$S_{p,i}$ the flux of the source  $i$  and $B_{p}$ the background, 
recorded during the pointing $p$. $D_{p}$ and $B_{p}$ are vectors of 19 elements.\\

 In the present work, we describe the background as  $ B_{p} = A_{p} U $ where  $A_{p}$ is 
  the normalisation coefficient per pointing and U the "uniformity map" of the SPI camera,
   derived from an empty field observation.
The system consists of $N_{d}$ (number of detectors) $\times$ $N_{p}$ (number of pointings) equations
 solved simultaneously by a chi-squared minimisation method.
The number of unknowns (free parameters) is $N_{p}$ $\times$ ($N_s$ + 1) (for the $N_s$ sources and 
the background fluxes)
but we can reduce them by realising that the time variability for the sources and the background is longer than a single pointing.\\ 
The  timescales  depend on the goal of the analysis. For the image reconstruction performed by the SPIROS 
detection algorithm (Skinner \& Connell 2003), source
 and background  fluxes are generally assumed to be constant during the whole set of observations.
 In SPIROS, the
source positions are extracted using an iterative source research technique, implemented 
for the coded mask telescope \textit{INTEGRAL} SPI. The fluxes determined in this way are thus rough mean fluxes,
the main goal being to extract source positions.   

To build the light curves of the sources detected in the field of view, we 
must choose the appropriate time scale for each component (sources and background). Concerning the background flux, the global count-rates registered with 
the Anti-Coincidence System (ACS) 
indicate  that the background flux is stable within
 each of the considered revolutions.
For the sources, it is important not to oversample the temporal variability as 
this increases the errors and provides no further scientific information.
 We have chosen a time scale for each source
mainly as a function  of its intensity, and of its a priori known or observed temporal behavior.
 The fainter sources have been considered to have a constant flux within a revolution. 
For the brightest sources, we test several values and choose the longest timescale
over which the source is found not to vary (see below). 
The system of $N_{d} \times N_{p}$ equations (1) is thus completed by a number of additional constraints reflecting
the non-variability of a given parameter ($S_{p,i}$ for a source or $A_{p}$ for
background) over its timescale ($\Delta t_i$).
The resolution of this system by the chi-squared minimisation method gives the light curves of
all components simultaneously.

The count spectra are constructed by solving a similar set of equations in a number of energy bands, and then
deconvolved using  the energy response matrix corresponding to each pointing (Sturner et al. 2003)
to get the photon spectra. We added a $3 \% $ systematic error to all spectral channels, and the spectra were
fitted from 23 to 600 keV in XSPEC.

Only pointings for which GX 339-4 was at a distance lower than 12 degrees
from the central axis were taken into account for the analysis.
 We excluded pointings affected by a solar flare or by 
exit/entry into the radiation belts.
We obtained 273 ks of useful data 
 with 140 pointings during the observation period covered by the revolutions 166, 167, 174  and 175 (see Table~\ref{tab:table1}).

\subsubsection{IBIS}
The Imager On Board \textit{INTEGRAL} (IBIS) is a coded mask telescope with a total field of view of
 29$^\circ\times$ 29$^\circ$ (down to 0 response) composed of a two layers detection plane.
 We use here only data from the upper layer, the \textit{INTEGRAL} Soft Gamma-Ray Imager
 (ISGRI, Lebrun et al. 2003). This detector  is sensitive
 from $\sim 13$ keV to about $1000$ keV. 

  Due to its low efficiency and the general faintness of the fluxes of sources  above
 $\sim$400 keV, it has generally a limited use above that energy.  

 IBIS/ISGRI has excellent imaging capibilities
 with an angular resolution of 12$\arcmin$ and a positional accuracy  
 around 1 $\arcmin$ depending on
 a given source signal to noise ratio. ISGRI also possesses spectral  
 capabilities, although with a much poorer
 resolution than SPI, with $\Delta$E/E $\sim$0.1 at 60 keV.

The data from IBIS/ISGRI were reduced in a manner strictly  
identical to the one reported in Rodriguez et al. (2006) which  
focuses on the same field with the exception that the last version of  
the Offline Scientific Analysis ({\tt{OSA}})
software (V 5.1) was used. To quickly summarize, we first produced  
images in 2 energy ranges (20-40 keV, and
40-80 keV), to identify the most active sources. We then extracted  
spectra from all the sources in this field
that had a detection significance higher than 7. We rebinned the  
original response matrix (${\tt{isgri\_rmf\_grp\_0016.fits}}$) 
to 63 channels before the extraction. The latest time  
dependent ancillary response file (${\tt{isgri\_arf\_rsp\_0013.fits}}$) was  
then associated to the file for the spectral fits.
We added 2$\%$ systematics to all spectral channels, and the spectra  
were fitted (together with those from the other instruments) from 25 to 200 keV.

\subsection{\textit{RXTE}}
We also analysed the public PCA+HEXTE (Bradt et al. 1993, Rothschild et al. 1998) data. Table~\ref{tab:table1}
 summarizes the set of \textit{RXTE}
observations performed contemporaneously with the \textit{INTEGRAL} data observation periods.
 
We limited the energy range from 3 to 25 keV, and 25 to 200 keV for the PCA and HEXTE data
respectively. The data reduction was done using the FTOOLS routines in the HEAsoft software package distributed by NASA's HEASARC (version 5.2).
We followed the steps of standard reduction as explained in the \textit{RXTE} Cook Book.  
The spectrum extraction was performed from data taken in "Standard 2" mode. The response matrix and 
the background spectra were created using FTOOLS 
programs. Background spectra were made using the latest "bright source" background model.

Only the proportional counter unit (PCU) number 2
of the PCA (Bradt et al. 1993) detector
has been used for the data extraction.
 We added 0.8\% up to 7 keV and 0.4\% above 7 keV as 
systematic error (for the details of how we estimated systematic
uncertainties, see Tomsick et al. 2001).

For HEXTE, we used the response matrix 
created by the FTOOLS, and applied 
the necessary dead time correction (Rothschild et al. 1998). The HEXTE
background is measured throughout the observation
by alternating between the source
and background fields every 32 s. The data from the background regions are merged. HEXTE channels were grouped by 2
for channels 16-31, by 4 for channels 32-59, by 10 for channels 60-99 
and by 64 for channels 100-227.

By fitting HEXTE data during the observation period of revolution 175 with a power-law plus cutoff model, we found a
$\chi ^2$ value of 2.15 (12 dof). All the fits including these HEXTE points are clearly degraded.
Therefore we suspect a problem for this particular data set (see $\chi ^2$ values in tables 4, 5 and 6). Thus we decided for this particular revolution, to perform the analysis with and without the HEXTE data.
We stress that including the HEXTE data vastly degrades the  
$\chi ^2$, thus we don't consider the corresponding results in the discussion.

\section{Results}

\subsection{Field of view around GX 339-4: flux extraction}

Several hard X-ray sources are present in the GX 339-4 field and have to be carefully taken
into account with appropriate variability time scales in the SPI analysis.

The sources  detected above 20 keV using SPIROS, for revolutions 166 + 167, are listed with their 
23-50 keV mean fluxes in Table~\ref{tab:table2}.

 As mentioned in Section 2.1, the errors increase with the number of sources and temporal bins.
So we make sure that each additional degree of freedom really improves
the $\chi ^2$. First, due to its highly variable behavior, the 4U1700-377  light curve
in the 23-200 keV energy range requires the use of the highest timescale resolution, 
ie one pointing  (whose duration is  about 1770 s during revolution 166, 2700 s  for the others).
Then  timescale variabilities of a 3 
(for 4U 1630-47, GX 354-0 and GX 340+0) and 2 (for OAO 1657-415) 
pointings duration  
were found to significantly improve the $\chi ^2$. 
We consider finally 6, 4 and 2  other sources, including GX 339-4, with a constant flux 
for  revolutions 166+167, 174 and 175 respectively.\\
The resulting 23-44 and 44-95 keV fluxes obtained with SPI 
 for GX 339-4 are presented in Table~\ref{tab:flux}. The quoted errors are at 1 $\sigma$ level.
It appears that the source flux increases by a factor 3 between revolutions 166+167 and revolutions 174 and  175.

When going to higher energies, the sources significances decrease (see Table~\ref{tab:table2})
 and we thus consider only 5 sources to  extract fluxes above 150 keV.
 
\subsection{Spectral evolution}

The GX 339-4 spectra corresponding to
each set of data detailed in Table~\ref{tab:table1} have been fitted with various
models available in the standard XSPEC 11.3.1 fitting package  (Arnaud 1996). 
In all cases,  we account for the interstellar absorption (PHABS in XSPEC) 
using a column density $N_H$  of $3.7 \times 10^{21}$ cm$^{-2}$ 
(Miller et al. 2006). 
In all fits, the 
 iron emission line was modelled by a narrow Gaussian fixed at an energy 
 of 6.4 keV. As
the
 width of the line is only weakly constrained it was fixed at 0.1 keV.
For all models, the inner disk inclination was fixed at $50^\circ$.
The data of revolutions 166 and 167 have been added to improve
 statistics, due to the low flux of the source and short duration of the revolution 167.

We used the simultaneous PCA (3-25 keV), 
HEXTE (25-200 keV), IBIS (25-200 keV) and SPI (23-600 keV) data for the spectral analysis.

\subsubsection{PEXRAV model}

First, we fitted all data with the PEXRAV model (Magdziarz \& Zdziarski 1995)
consisting of a power-law with a high energy cut-off and reflection from neutral medium.

Fixing the  reflection fraction  to $\Omega = 0$ 
 provides  a marginally acceptable fit for revolution 166+167 (reduced $\chi ^2$ of 1.11)
and a very poor description of the data for 
revolutions 174 and 175 (reduced $\chi ^2$ of 2.12 and 2.86 respectively).
The pattern of residuals is
 characteristic of the presence of 
Compton reflection.
Adding a reflection component improves the fit dramatically, reducing
the $\chi ^2$ to 0.82, 0.91 and 1.09 for revolutions 166+167, 174 and 175
respectively (see Table~\ref{tab:table7bis}).

The addition of a Compton reflection gives a better fit at
a high significance during all those
observations, with a probability
$\la 10^{-13}$ that adding this component is not required by the data (as obtained 
using the F-test).

We see in Table~\ref{tab:table7bis} that the reflection fraction increases and reaches a value
of $\sim$ 0.5 when the flux is high.
The  photon index is around 1.6-1.7 as usually observed, 
but the energy cut-off is rather high
(300-400 keV).  
To test the reality of the cut-off, we perform fits with the same model
but fixing Ec to 2 MeV (i.e. outside our energy range).
The fit with a free high energy cut-off is significantly better 
in all cases with F-test probability of $10 ^{-12}$, $3 \times 10 ^{-17}$  
and $2 \times 10 ^{-12}$ for revolutions 166+167, 174 and 175 respectively.
 We attempted to fit with ionized reflection using the PEXRIV model but
the best fit  ionization parameter tends to zero.

\subsubsection{Physical models based on Comptonization}  
To go more deeply into the understanding of the source behavior, we have used
more sophisticated models, based on the Comptonization process as it is thought to be the 
main mechanism able to produce the emission observed in our energy domain.
  
\subsubsubsection{COMPPS model}   
 First, we modelled the X- and $\gamma$-ray spectrum with the thermal  
 Comptonization model COMPPS (Poutanen \& Svensson 1996). 
Blackbody seed photons are injected 
into a spherical corona of uniform
  optical depth $\tau$,  and temperature kT where they are Comptonized.
  The temperature of the blackbody component cannot be constrained and was 
  frozen to 390 eV (Miller et al. 2006). 
 
 A fraction of the hard X-rays is scattered back into the disk where
  it is reflected. 
 We consider the case of reflection from cold, neutral material with solar 
abundances.

As seen from Table~\ref{tab:table8bis}, the temperature and the optical
depth can be considered as constant 
as well as the equivalent width which was found to a value 
of about 90 eV. Only the reflection fraction varies,  increasing from 0.2 up to 0.4, between 
the first (low intensity)
and the 2 last observations (brighter by a factor of 3).\\

\begin{figure*}%[!t]
\begin{center}
\includegraphics[scale=0.4]{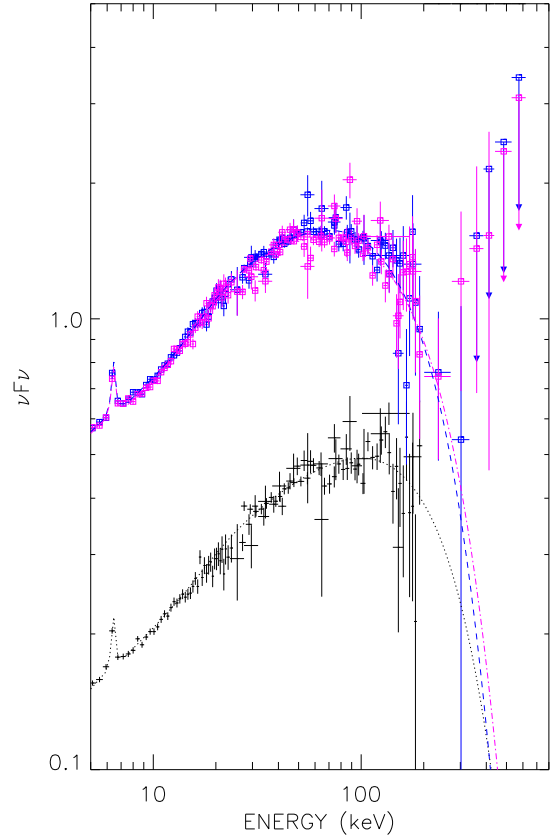}
\end{center}
\caption{Deconvolved spectra of GX 339-4 with simultaneous PCA (3.-25 keV),
 HEXTE (25-200 keV), SPI (23-600 keV) and ISGRI (25-518 keV) data,
for revolutions 166+167 (black  points), 174 (blue squares) and 175 (magenta squares). 
The lines correspond to the best fits with the EQPAIR thermal model (see table 6).}
\label{fig:fig3a}
\end{figure*}

\begin{figure*}%[!t]
\begin{center}
\includegraphics[scale=0.4]{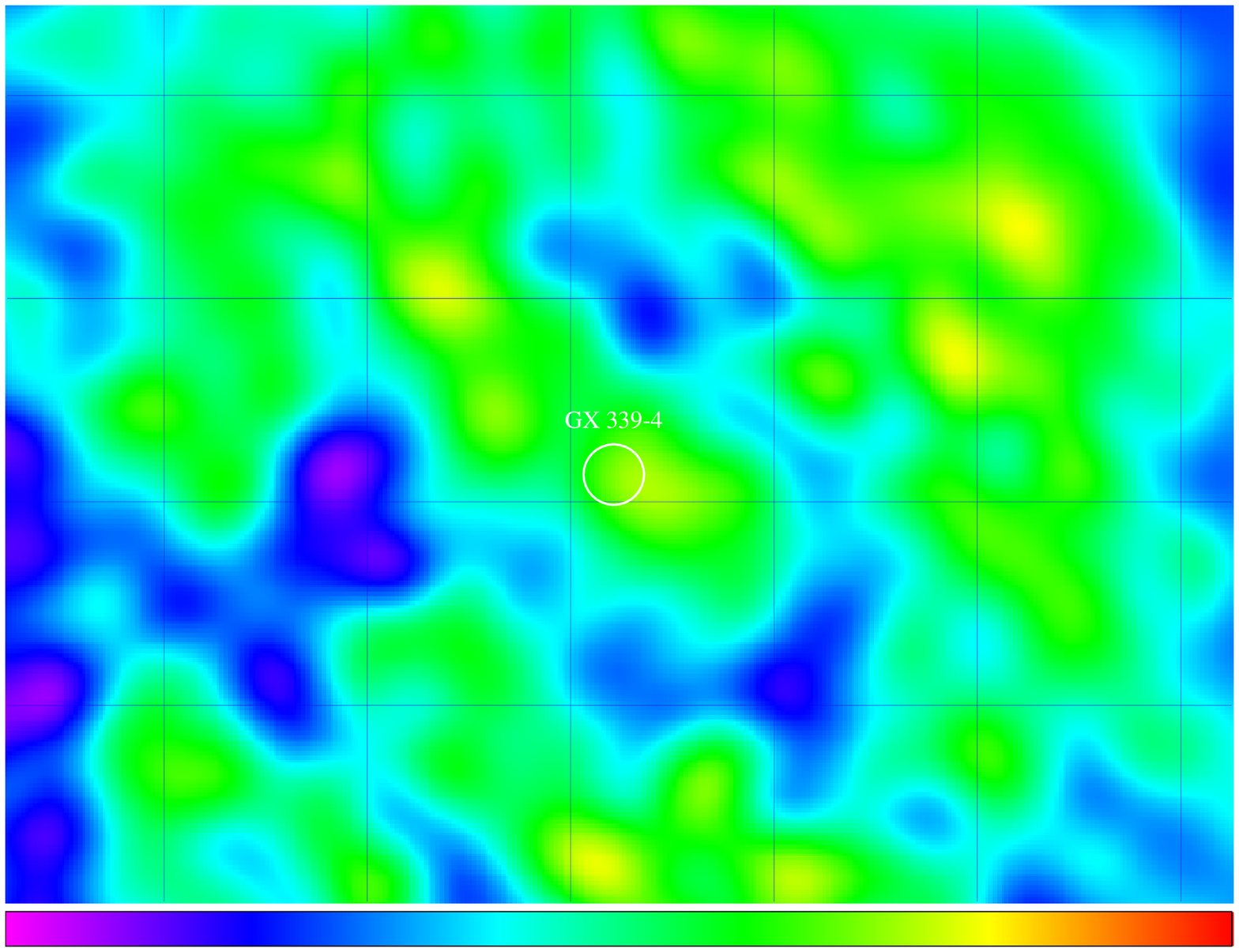}
\includegraphics[scale=0.4]{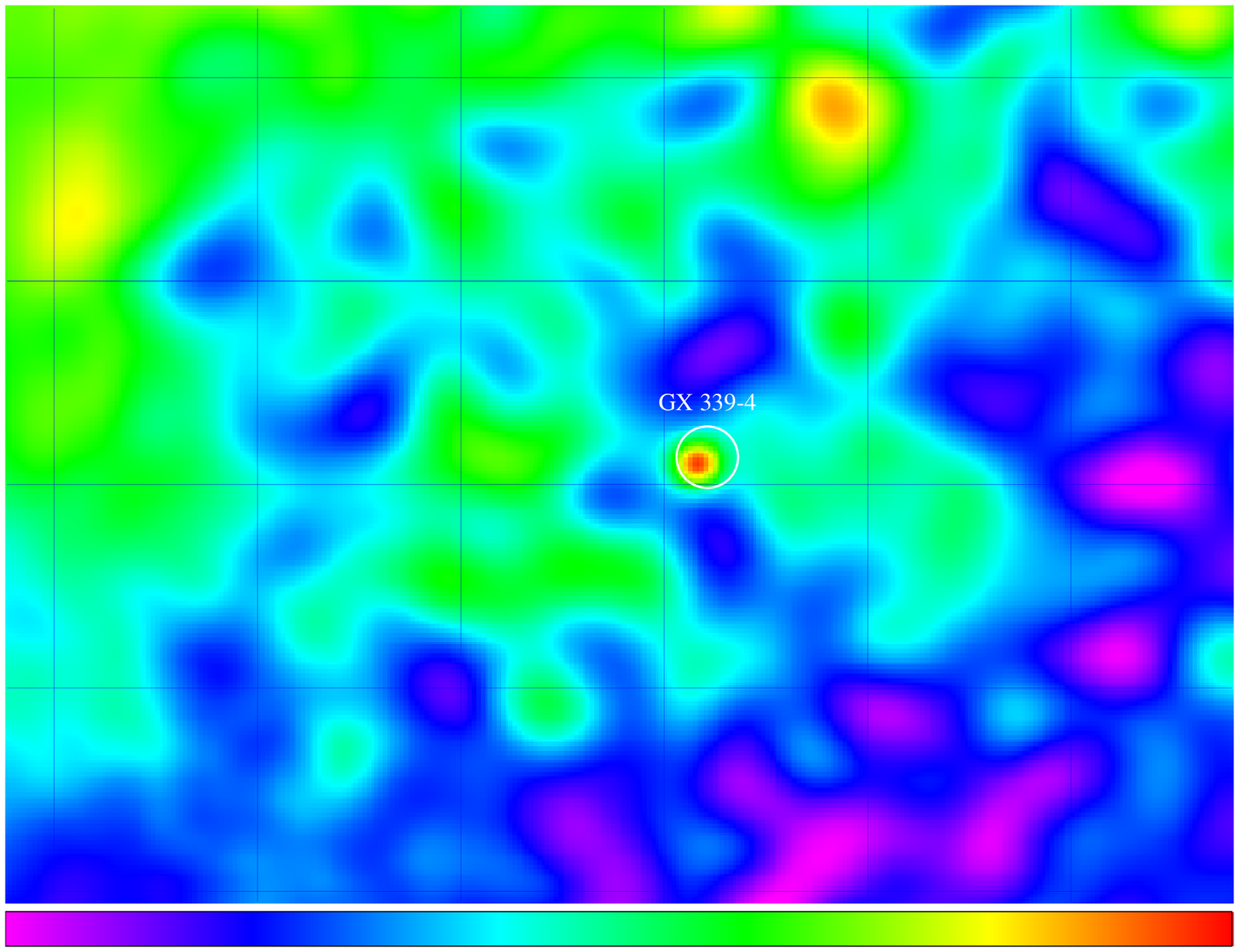}
\end{center}
\caption{Significance maps obtained with SPIROS in the 200-437 keV 
energy range; the color scale is from -3$\sigma$ to 4.5$\sigma$; grid spacing is 5$^\circ$.
Left: revolution 174, the significance of the
GX 339-4 flux is below 1.5 $\sigma$. Right: revolution 175,
GX 339-4 is the only  significant source 
of the field of view detected with a flux of 193.4 $\pm$ 46.6 mCrab (4.2 $\sigma$ ).}
\label{fig:Images-HE}
\end{figure*}

We see from Figure \ref{fig:fig3a} that the global shape of
the spectrum is unchanged between observations  
174 and 175 as well as the flux of the source (see Table~\ref{tab:table8bis}). But in the latter, an excess of emission relatively to 
the model appears above 200 keV.
This is unexpected from thermal model and we thus investigated this point more deeply.\\
We first reconstructed the image in the 200-437 keV energy band
for both revolutions 174 and 175 (Figure~\ref{fig:Images-HE}). While the image of revolution 174 doesn't reveal
any significant feature, the highest excess detected by SPIROS for revolution 175 coincides with the GX 339 position.
Note that the distribution and the level of the residuals in both images follow the expected ones, thus
allowing to be confident in the data reduction process. We can assess that 
all statistical tests show a low level of systematics. Thus the measured significance of the emission
(4.6 $\sigma$) has not to be corrected.\\

\begin{figure*}%[!t]
\begin{center}

\includegraphics[scale=0.39]{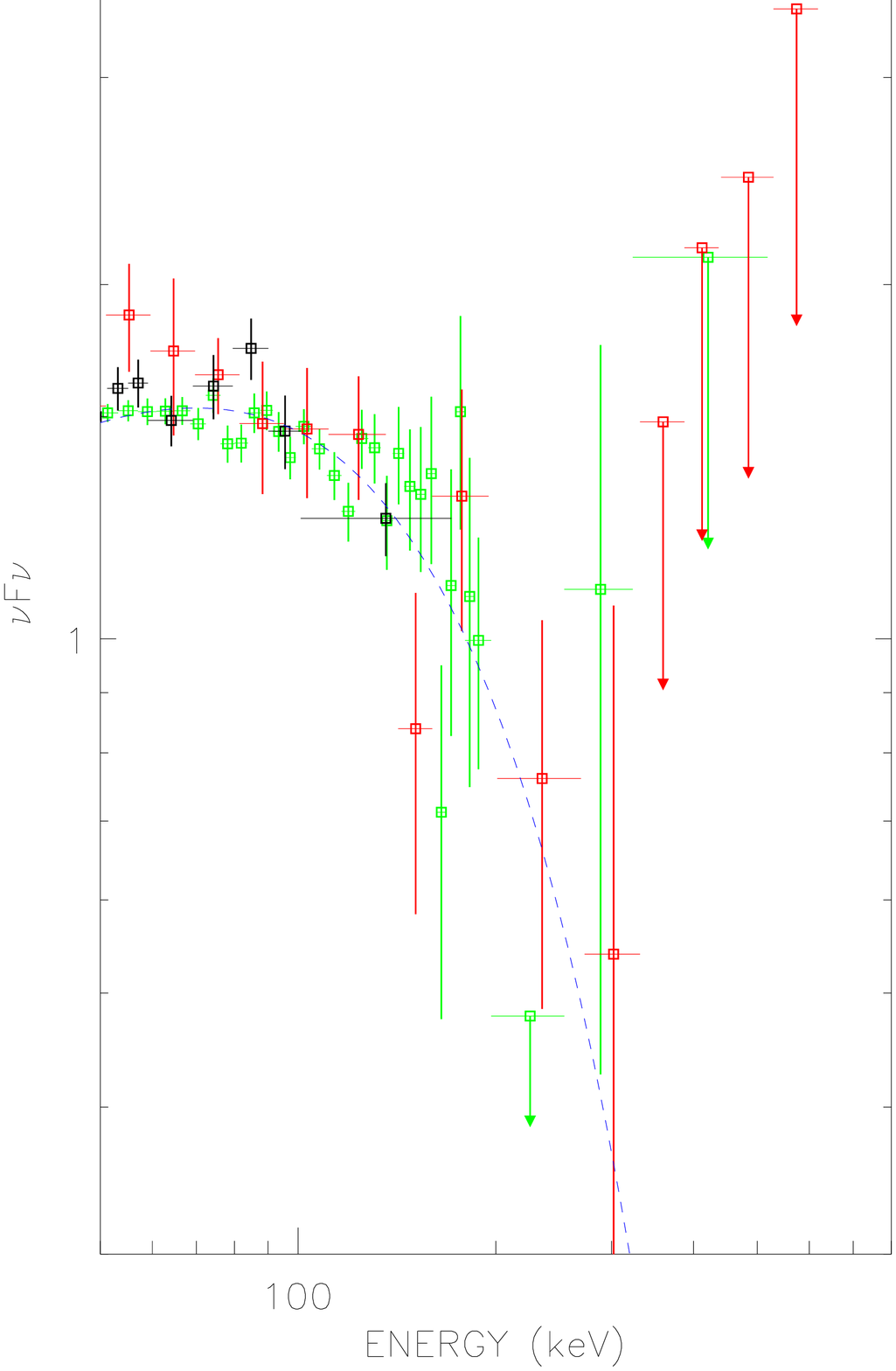}
\includegraphics[scale=0.39]{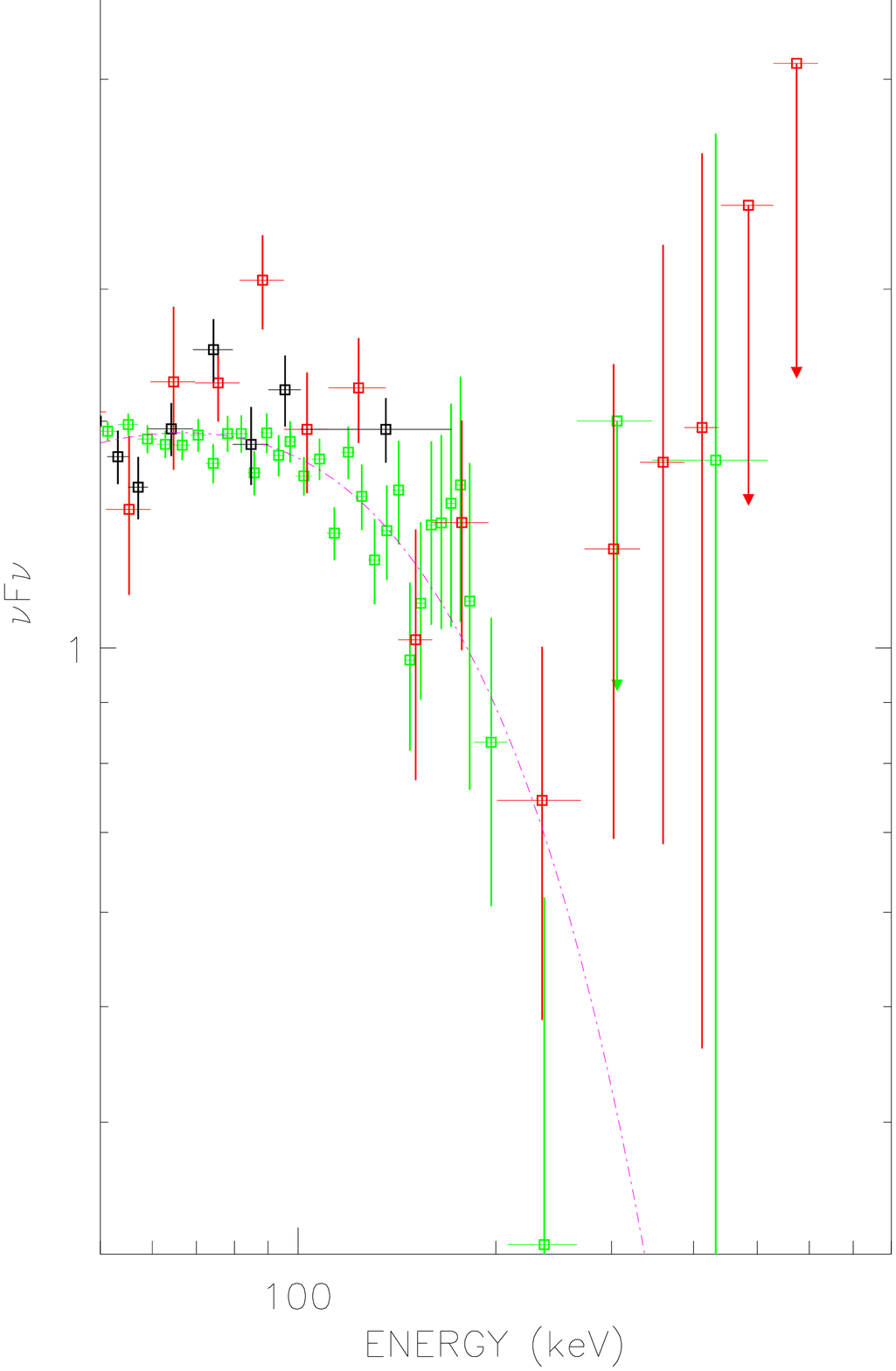}
\end{center}
\caption{High energy part of the GX 339-4 deconvolved spectra for revolution 174 (left) and for
revolution 175 (right) with  HEXTE (black points), SPI (red points) and ISGRI (green points) data. 
The line corresponds to the best fit with the EQPAIR thermal model (see table 6).} 
\label{fig:HE-SPECTRES}
\end{figure*}

\begin{figure*}%[!t]
\begin{center}
\includegraphics[scale=0.4]{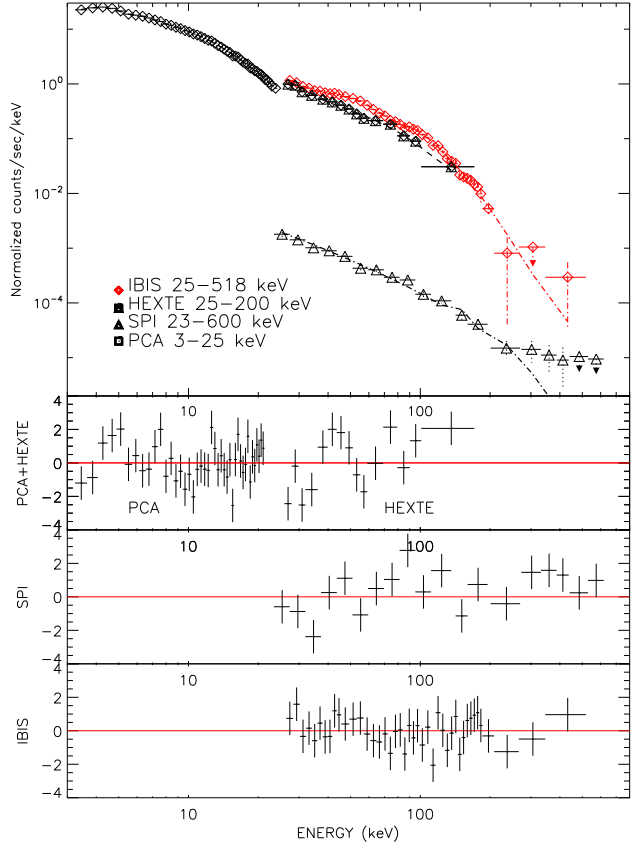}
\end{center}
\caption{Spectra of GX 339-4 with simultaneous PCA, HEXTE, ISGRI and SPI data
for revolution 175. The COMPPS model used is described in Table 5. The residuals 
obtained for each instrument are presented in the bottom panels.}  
\label{fig:175.all}
\end{figure*}

Figure \ref{fig:HE-SPECTRES} presents the high energy part of the source spectrum
for revolutions 174 and 175, while in Figure \ref{fig:175.all} are displayed the data for all instruments,
together with the corresponding residuals relatively to the COMPPS model. 
They show that SPI and IBIS data are in agreement even if the IBIS points above
200 keV are not significant.
The IBIS measurment in the 200-437 keV energy band gives an 3$\sigma$ upper limit of 223 mCrab, to compare to the
SPI measurement of 218 ($\pm$ 47) mCrab. 

We have thus used only the SPI data to quantify this excess relatively to the thermal emission.
The shape of the SPI spectrum during the revolution 175 has been fitted using the
COMPPS model. The best fit parameters are  an electron temperature kT
of 44 keV and an optical depth of 4.4 (with a reduced $\chi ^2$ of 
1.35). 
We found that the addition of a simple power-law 
leads to an improvement of the fit of the SPI data significant  at the 90 \% level
 according to a F-test , with a best fit photon index of 1.07.

\subsubsubsection{EQPAIR model}  
 We  finally applied the hybrid   thermal/nonthermal Comptonization  model EQPAIR (Coppi, 1999).
  This model assumes a spherical plasma cloud with isotropic and
  homogeneous distribution of electrons, positrons
   and soft seed photons  within the plasma. 
  The properties of the plasma depend on its compactness 
  $l=L \sigma _T / R m_e c^3$. $L$ is the power of the source, $R$ the radius of the 
  sphere which is assumed  to be $10 ^7$ cm and $\sigma _T$ is the Thomson cross-section.
  We use a hard compactness $l_h$ which corresponds to the power supplied to the 
  electrons, and a soft compactness $l_s$, corresponding to the power
  supplied in the form of soft seed photons.
The amount of heating of the Comptonizing medium is specified 
through the 
ratio of the compactnesses of the Comptonizing medium $l_h$, and of the
seed photon distribution $l_h$/$l_s$ with $l_{s}$ fixed to 1. 
The seed photon blackbody temperature kT$_{seed}$ is frozen 
to 390 eV. The reflection
component is modelled as in COMPPS and the reflection fraction $\Omega /2 \pi$ 
obtained from the fit corresponds to the unscattered part of the Compton
reflection component.
The total optical 
depth, ($\tau _{tot}$), is the sum of the optical depth of e$^+$e$^-$ pairs ($\tau _{e^+ e^-}$),
and of e$^-$ coming from ionization of atoms ($\tau _{es}$).
The fitted parameters are $l_h$/$l_s$ (which is related to the coronal temperature)
and the ionization electron optical depth, $\tau _{es}$.\\ 
 In order to consider the case of a hybrid plasma, 
 the model contains an additional parameter: the ratio $l_{nth}$/$l_{th}$ where $l_{nth}$ is
 the compactness corresponding
 to the non-thermal and $l_{th}$ to the thermal part of the e$^+$e$^-$  distribution.
 The rate at which non-thermal electrons appear in the source is assumed to be a 
 power-law, $\gamma ^{-\Gamma _{inj}}$ between the Lorentz factor $\gamma _{min} = 1.3$
 and $\gamma _{max} = 1000$. $\Gamma _{inj}$ is assumed to be equal to 2.
We combined EQPAIR with a gaussian iron line centered and fixed 
at 6.4 keV with a width of 0.1 keV. 
Table~\ref{tab:table20} shows the evolution of the best fit parameters.
\\
We first fixed  the compactness of the non-thermal electrons 
to zero
in order to consider the case of a purely thermal plasma.
Spectra extracted during all the observations 
are well fitted with $\chi ^2$ $\simeq 0.8-1.1$. 
 The hard-to-soft compactness ratio is in the range 
 of 5-6 yielding a coronal temperature of about 65 keV, given the fitted
optical depth of 1.6 in revolutions 174-175 while the kT (98 keV) and $\tau _{tot}$ (1.1) parameters of revolutions 
166+167 are very different. 
 Part of the difference between the kT and $\tau$ best fit values can be explained by the fact that the Comptonization model
suffers 
from a degeneracy in their determination due to low statistic. 
Indeed the spectral slope depends only on the
Compton parameter
 $y$ = $4\times kT\times \tau /m_{e}c^2$ which does not vary a lot along the observations (see Table~\ref{tab:table20}).
In fact, if we freeze 
the optical depth in revolution 166+167 to the value obtained in revolution 174 
we obtain an acceptable $\chi ^2$ value and
the difference in temperature between both revolutions decreases.
 Nevertheless the $y$ parameter seems to be larger in revolution 166+167 which is confirmed by a 
 significanlty higher $l_h$/$l_s$.

The reflection component fraction is similar to the value found 
with COMPPS model and the iron line equivalent widths are still about 90 eV.

In a second step, the compactness of the non-thermal distribution has been freed in order to consider
the case of an hybrid plasma.\\
For revolution 174, this leads to a thicker ($\tau _{tot}$ $\simeq$ 2.0) 
and cooler (kT$_e$ $\simeq$ 50 keV) solution, relatively to the thermal
case, with both solutions equivalent from a statistical point of view.

Moreover, determing the confidence range of the $l_{nth}$/$l_{th}$ parameter, we found
it unconstrained, forbidding any conclusion.

The situation seems different for the revolution 175 as the introduction of a non-thermal component allows
a slight improvement of the $\chi ^2$ (with a FTEST probability of $10^{-2}$) even though the non-thermal fraction is not really constrained. 
However, the parameters are the same as in the thermal case except the hard compactness which slightly increases.

\section{Discussion}

\subsection{Comparison with previous observations}
Broadband  X and $\gamma$-ray spectrum of GX 339-4 using data collected from
SPI, IBIS, HEXTE and PCA instruments allowed us to follow the source behavior 
during the low/hard state corresponding to the rising phase of its 2004 outburst.
During this period we observed a flux increase by a factor 3 in the whole energy
domain (3 - 200 keV) without any major change in the spectral shape (see Figure \ref{fig:fig3a}). This is similar to results from previous outbursts reported
by Zdziardski et al. (2004, Figure 4) where, during the LHS, in the rising phase, the
 source flux increases at constant spectral
slope.

\subsubsection{High energy cut-off and the reflection component}
We showed that reflector and high energy cutoff in the primary emission
component were required to describe the spectral shape during
the LHS, as typically observed for GX 339-4.
Simultaneous  fits from PCA, HEXTE, IBIS and SPI data  give photon indexes $\Gamma$  in the range of 1.6-1.7 with
cut-off energies around 300-400 keV.
The amplitude of the reflection component 
reaches similar values to those reported by
Revnivtsev et al. (2001), who modelled the low$/$hard state of GX 339-4
during its oubursts of 1996-1997 as observed by PCA. They found
photon indexes $\Gamma$ in the range  of 1.7-1.9 and a
reflection amplitude of 0.3-0.5.
Moreover the reflection fraction
increases with the photon index of the power-law and flux in a way similar to
the correlation already observed by
Nowak et al. (2002) for GX 339-4 and in a large sample of Seyfert and X-ray binaries by 
Zdziarski et al. (1999, 2003).
Zdziarski et al. (1999)
 have interpreted the $\Omega$-$\Gamma$ correlation as being due to
feedback in an inner hot (thermal) accretion flow surrounded by an overlapping
cold disc. Then, the closer to the central black hole the cold disc
extends, the more cooling of the hot
plasma by blackbody photons (indicated by the
decrease of the electron temperature kT fitted by the COMPPS model)
and the softer the spectrum (as shown from the powerlaw slope $\Gamma$
fitted by PEXRAV model).

\subsubsection{Comptonization parameters }
We then introduced the physical models COMPPS and EQPAIR to describe the
GX 339-4 observations with 
Comptonization process. From a statistical point of view, both models give acceptable solutions.
  Data from revolutions 
 166+167 and 174 are well fitted with Comptonization model in a purely 
 thermal case, while the last revolution suggests the presence of a non-thermal component.

Wardzinski et al. (2002) modelled the
ouburst of September 1991 observed by Ginga$/$LAC and CGRO$/$OSSE
as well as the outbursts of 1996 and 1997 observed by
\textit{RXTE} PCA+HEXTE with COMPPS model. 
The Comptonization spectrum was caracterised by Thomson optical depthes
$\tau$ of 2.4-3.0 and electron temperature of 60 keV.
The reflection component was found to be moderately ionized with amplitudes
 ranging from 0.2 to 0.4.
 We have tested 
 the ionized case, but the ionization parameter
 tends to zero. We have thus fixed it to zero and in fact deduced 
 very similar best fit
 parameters.

The EQPAIR thermal model has been used by
Nowak et al. (2002) to  interpret the spectra obtained during the 
low$/$hard state observations of the GX 339-4 outbursts in 1997 and 1999
 by PCA+HEXTE.
With a fitted seed photon temperature of 30-100 eV, 
they found  coronal compactnesses l$_c$  ranging
from
5 to 14 and electron optical depthes $\tau _{es}$  from 0.01 to 0.6, yielding
to total optical depthes $\tau _{tot}$ between 0.1 and 1 and kT$_e$ around 200 keV.
The reflection fractions 
are between 0.1-0.5 and the equivalent widths of the iron line range from 80 to 240~eV.
We find similar values for the reflection fraction, iron line width and coronal
compactness, but thicker ($\tau _{tot}$ around 1.6-1.7) and colder 
(kT$_e$ around 65 keV) plasma parameters.

\subsection{The high energy excess}

An interesting feature appears in the SPI data at revolution 175
as the emission extends beyond the thermal cut-off. 
Such an excess above 200 keV has already been observed 
by the OSSE observatory (Johnson et al. 1993)
in the low/hard state of 
the GX 339-4 outburst event in September 1991. 
Assuming a distance of 6 kpc, the luminosity above 200 keV 
was $11.3 \times 10^{36}$ erg s$^{-1}$ (Johnson et al. 1993)
with a 35-300 keV luminosity L$_{35-300~keV}$ of 2.5 
$\times 10^{37}$ erg s$^{-1}$.
These values are comparable to what we found  for 
 the excess observed during revolution 175 with a luminosity above 200 keV of 
 $7.2 \pm 1.6 \times 10^{36}$ erg s$^{-1}$ and L$_{35-300~keV}$
$\simeq$ 1.54 $\times 10^{37}$ erg s$^{-1}$.
More recently, a similar feature has been reported in one \textit{RXTE} observation
 (Nowak et al. 2002), where the HEXTE spectrum exhibits a hardening above $\sim$ 100 keV.\\
Similar behavior is observed for Galactic black hole transients in the LHS during the ourburst
decays, especially after the detection of compact jets (Kalemci et al. 2006; Kalemci et al. 2005).

The flux extension above a thermal Comptonization component can be explained by a second Comptonization region/population or, more or 
less equivalently,   spatial (gradient) and/or temporal 
variations in the plasma parameters. As our fitting models assume them to be 
homogeneous and constant during one observation, such variations can produce
deviation relatively to the predicted spectrum.
Another possibility could be the presence of a large annihilation
line which would produce a bump extending down to 250-300 keV.
Alternatively, some models include the presence of a non-thermal 
electron population in the plasma.
The  high energy power-law tail observed for Cyg X-1
in its LHS has been modelled with
a hybrid Comptonization, by injecting
an electron population consisting of Maxwellian distribution coupled to a power-law with an index of  4.5
(McConnell et al. 2000).
The presence of a high-energy electron tail has also been proposed by 
Wardzinski et al. (2002) for GX 339-4  simultaneous observations
of PCA+HEXTE and CGRO/OSSE during the
outburst of 1997, even though COMPSS model fits the data
reasonably well.
They found that the introduction of  a non-thermal fraction of 
electrons 
improves the $\chi ^2$, and corresponds to 34 \%
 of non-thermal emission. 
\\

The EQPAIR hybrid plasma allows a marginally better description  of the spectrum 
during revolution 175. 
Adding a non-thermal parameter l$_{nth}$ 
to the pure thermal plasma case gives an FTEST value of about $10^{-2}$.
However, the l$_{nth} /$ l$_{th}$ fraction ($\simeq$ 0.28) is  similar to the value found for the LHS of  GX 339-4 by Wardzinski 
et al. (2002), even though theses values are in fact not constrained.

In Galactic black holes, non-thermal emission is generally associated with the soft state
(Grove 1999, Ling et al. 1994) or
state transitions (see eg. Cadolle-Bel et al. 2006, Malzac et al. 2006),
 where power-law tails are 
observed, and attributed to Comptonization of  soft photons by accelerated electrons.
During the transition from the low/hard to the high/soft state on 
August 15 2003, Belloni et al. (2006) observed that the high energy cut-off 
increases/disappears.\\ 
In the LHS, the hard X/$\gamma$ ray emission comes from a 
 thermalised electron population.
The presence in the low/hard state of a significant non-thermal emission above the thermal cut-off 
can however be interpreted in terms of an additional component.
Contrary to the soft state where power-law tails are commonly observed, this characteristic is much rarer
 in the LHS. In our observations, this feature is detected only in one revolution,
nevertheless the statistic is not high enough to conclude on its variability. Moreover, as the low energy part of the spectrum is not affected (the spectral shape  is similar up to $\simeq$ 150 keV in revolutions  174 and 175),
this may suggest that the involved phenomenon varies 
independently of the thermal Comptonization process and could come from another location such as jets or active regions.
%The appearance of a high energy excess related to the jet itself has already been modelled by some authors 
Jet can easily produce hard X-ray emission via synchrotron radiation in addition to inverse Compton up-scattering (see e.g. Markoff et al. 2003).

\section{Conclusions} 
We investigated the high energy spectral characteristics of GX 339-4 
in the low/hard state (rising phase of the outburst), using observations from \textit{INTEGRAL} SPI+IBIS and 
\textit{RXTE} PCA+HEXTE. 
The plasma parameters deduced from thermal Comptonization models are identical to those obtained during previous observations performed when the
source was in a similar spectral state, with a plasma temperature kT around 65 keV, 
an optical depth $\tau \simeq 1.5-2.5$ and a reflection factor lower than 1 (0.2 - 0.4), supporting an anisotropic
primary emission or a large value of the disk internal radius.
An emission in excess relatively to the thermal cutoff has been detected 
during one revolution, while absent 3 days before.
The corresponding flux (typically  emission above 200 keV) is not highly significant
with only 4.6 $\sigma$, but can be compared to  features reported by OSSE/CGRO
and HEXTE in GX 339-4 or Cyg X-1. This can be an evidence that an 
additional component is sometimes present in the spectrum, unless variabilities
(spatial and/or temporal) of (kT, $\tau$) plasma parameters can mimic such spectral evolution affecting only the 
energy domain above 200 keV.

\acknowledgments
%!!!!!!!!!!!!!!!!!!!!!!!!!!!!!!!!!!!!!!!
The SPI project has been completed under the responsibility and leadership
of the CNES. We are grateful to ASI, CEA, DLR, ESA, INTA, NASA and OSTC
 for support. \\
Specific softwares used for this work have been developed by L. Bouchet.
E. K. acknowledges partial support of TUBITAK and Marie Curie International Reintegration Grant MIRG-CT-2005-017203.
The authors are grateful to the anonymous referee for its very fruitful comments that allowed us to improve the quality of this paper.
%!!!!!!!!!!!!!!!!!!!!!!!!!!!!!!!

\clearpage

\input{tab1}

\input{tab2}

\input{tab3}

\input{tab4}

\input{tab5}

\input{tab6}

\end{document}

%% file: tab1.tex
\begin{table}[htp!]
\centering
\begin{tabular}{lcccccccc} 
\hline
Rev.  &INT$_{start}$ &INT$_{stop}$& $\Delta$t$_{sp}$ (ks)&ID&RX$_{start}$& RX$_{stop}$&  Exp.(ks)\\
\hline
\hline
%GCDEa   &52674.16   &52724.35    &   252           &  -             &     -              &      -              & -        \\
%GCDEb   &52859.56   &52915.94    &   555              &  -             &     -              &      -              & -    \\
166    &53057.07   &53059.45    &   152             & 90109-01-01-00 &  53058.99          &    53059.17         & 16.1          \\
167    &53061.62   &53062.22    &   40               & 80132-01-07-00 &  53061.75          &    53061.86         & 9.3           \\
174    &53080.99   &53081.67    &   38              & 90118-01-06-00 &  53081.51          &    53081.54         & 2.8           \\
175    &53084.77   &53085.45    &   43               & 80102-04-66-00 &  53084.46          &    53084.49         & 2.5           \\
%226    &53236.53   &53237.74    &   72               & 90704-02-01-00 &  53237.32          &    53237.49         & 0.7          \\
%232    &53254.89   &53256.95    &   43               & 60705-01-73-00 &  53254.77          &    53254.78         & 1.2           \\
% 233    &53259.01   &53259.46    &   27               & -              &  -                 &   -                 & -           \\
\hline
\end{tabular}
\caption{The INTEGRAL observations of GX 339-4.
For each INTEGRAL revolution (Rev.), we give the beginning 
INT$_{start}$  and the end INT$_{stop}$ 
of the INTEGRAL observations in MJD.
$\Delta$t$_{sp}$ is the useful duration for SPI observations. 
ID is the identification program number of RXTE observations. RX$_{start}$ and RX$_{stop}$ are
the beginning and the end of RXTE observations taken (quasi-) simultaneously with INTEGRAL observations.
Exp. is the exposure time for PCA.} 
%\vspace*{-0.3 cm} 
\label{tab:table1}
\end{table}

%% file: tab2.tex
\begin{table}[htp!]
\centering
\begin{tabular}{lccccc} 
\hline
Source & $\Phi$ (23-50)  &$\sigma$ (23-50) &$\sigma$ (50-95)&$\sigma$ (95-195)\\
\hline
\hline
4U 1700-377   &  233.5$\pm$ 6.1&38.3& 11.9&5.3\\
4U 1630-47     &   85.6$\pm$  2.1&40.1&10.6&7.5\\
GX 339-4      &  44.2 $\pm$ 2.3&19.6& 10.5&8.5\\
IGR J16316-4028    &  21.9 $\pm$ 2.8 &7.9&  -&-\\
OAO 1657-415   & 26.7 $\pm$ 1.9 &13.6&-&-\\
4U 1636-536  & 44.4 $\pm$ 2.5 & 17.6&-&-\\
H 1705-440    &    68.2 $\pm$ 2.2& 31.0&-&-\\
GX 354-0    &   150.2 $\pm$ 5.8&25.9& 6.6&-\\
GX 340+0    &31.3 $\pm$ 2.1    &14.9& -   &-\\
1E 1740.7-2942    &   62.6 $\pm$ 15.9&3.9& 2.2&2.2\\
4U 1625-33         &27.5 $\pm$ 4.4   &6.3& 2.9&2.2\\
\hline
\end{tabular}
\caption{Sources detected by SPI in the field of view of GX 339-4
during revolution 166+167: $\Phi$ is the flux in the 23-50 keV energy range and
 $\sigma$, the significance  in the 23-50 keV, 50-95 keV and 95-195 keV energy ranges.} 
%\vspace*{-0.3 cm} 
\label{tab:table2}
\end{table}

%% file: tab3.tex
\begin{table}[htp!]
\centering
\begin{tabular}{lccc} 
\hline
Revolution & $\Phi _{23-44}$ & $\Phi _{44-95}$ \\
\hline
\hline
%GCDEa   &  13 $\pm$ 7&15 $\pm$ 4\\
%GCDEb   &  6 $\pm$ 2&7$\pm$ 2\\
166+167  &  42 $\pm$ 2&59 $\pm$ 4\\
174 &  156 $\pm$ 6&211 $\pm$ 10\\
175 &  152 $\pm$ 6&201 $\pm$ 10\\
\hline
\end{tabular}
\caption{Flux of GX 339-4 observed by SPI, for datasets introduced in Table 1, in the 23-44 keV and 44-95 keV energy ranges (expressed in mCrab).} 
%\vspace*{-0.3 cm} 
\label{tab:flux}
\end{table}

%% file: tab4.tex
\begin{table}[ht]
\renewcommand{\arraystretch}{0.01}
%\centering
\begin{tabular}{lccccccccc} 
\hline
rev   &   $\Gamma$       &  E$_{c}$     &$\Omega$/${2\pi}$      & W$_{Fe}$ &$\chi ^2$(dof)&F-test\\	
      &                  &  keV         &         & eV &              &    \\
\hline
\hline
166+167	&$1.58^{+0.02}_{-0.02}$ &$365^{+100}_{-67}$&$0.22^{+0.06}_{-0.06}$&$88^{+33}_{-29}$ &0.82(172)&\\
166+167	&$1.65^{+0.01}_{-0.01}$ &2000 (fr)          &$0.35^{+0.05}_{-0.05}$&$79^{+35}_{-30}$&1.10(173)&$1.0 \times 10^{-12}$\\
174	&$1.64^{+0.02}_{-0.02}$ &$293^{+60}_{-41}$&$0.43^{+0.08}_{-0.07}$& $94^{+43}_{-36}$&0.91(114)&\\
174	&$1.77^{+0.01}_{-0.01}$ &2000 (fr) &$0.80^{+0.06}_{-0.07}$& $94^{+34}_{-34}$&1.68(115)&$2.7 \times 10^{-17}$\\
175 	&$1.67^{+0.02}_{-0.02}$ &$325^{+68}_{-50}$&$0.50^{+0.08}_{-0.07}$& $77^{+38}_{-36}$&1.09(100)&\\
175	&$1.78^{+0.01}_{-0.01}$ &2000 (fr) &$0.85^{+0.07}_{-0.07}$& $82^{+40}_{-41}$ &1.78(101)&$1.53 \times 10^{-12}$\\
175*	&$1.69^{+0.02}_{-0.02}$ &$402^{+103}_{-71}$&$0.52^{+0.08}_{-0.08}$& $77^{+38}_{-36}$&1.60(114)&\\
175*	&$1.79^{+0.01}_{-0.01}$ &2000 (fr) &$0.82^{+0.06}_{-0.05}$& $72^{+34}_{-36}$ &2.02(115)&$1.94 \times 10^{-7}$\\
\hline
\end{tabular}
\caption{PCA, HEXTE, SPI and ISGRI data fitted simultaneously 
using the XSPEC multicomponent model PHABS*(PEXRAV+GAUSSIAN).
$\Gamma$ is the photon index and E$_c$ the energy cut-off.
Gaussian line was fixed at an energy of 6.4 keV
with a width fixed to 0.1 keV. W$_{Fe}$ is the equivalent width. $\Omega$/${2\pi}$ is the reflection fraction.
The F-test is calculated between models with free cut-off and no (ie fixed) 
cut-off. We show for the dataset of revolution 175, the fit parameters in the case for which 
the HEXTE data have been included (175*).}
\vspace*{-0.3 cm} 
\label{tab:table7bis}
\end{table}

%% file: tab5.tex
\begin{table}[ht]
\renewcommand{\arraystretch}{0.01}
%\centering
\begin{tabular}{lcccccccc} 
\hline
\hline
rev   &  kT  & $\tau$& W$_{Fe}$  & $\Omega$/${2\pi}$& L$_{2-600}$                &y  &$\chi ^2$(dof)\\	
      &  keV &       &keV               &                       &$\times 10^{36}$ erg s$^{-1}$&  &               \\
\hline
\hline
%166+167	&$45^{+8}_{-7}$&$4.04^{+0.79}_{-1.03}$ 87 &$0.10^{+0.07}_{-0.06}$&$12.8^{+1.0}_{-1.0}$ &1.42&1.00(134)\\
%166+167	&60 fr&$3.02^{+0.08}_{-0.11}$ &87 &$0.12^{+0.05}_{-0.05}$&$12.8^{+1.0}_{-1.0}$          &1.42&1.02(135)\\
166+167	&$72^{+21}_{-18}$&$2.40^{+0.72}_{-0.64}$ &$87^{+28}_{-46}$ &$0.23^{+0.06}_{-0.06}$&$12.8^{+1.0}_{-1.0}$          &1.35&0.88(172)\\
174	&$64^{+10}_{-7}$&$2.43^{+0.48}_{-0.53}$ &$91^{+31}_{-40}$ &$0.37^{+0.06}_{-0.05}$&$35.5^{+0.1}_{-0.2}$&1.22&0.82(113)\\
175	&$61^{+12}_{-8}$&$2.54^{+0.46}_{-0.48}$ &$90^{+30}_{-34}$&$0.43^{+0.06}_{-0.06}$&$34.9^{+0.1}_{-0.3}$ &1.22&1.28(98)\\
175*	&$59^{+11}_{-7}$&$2.64^{+0.44}_{-0.46}$ &$91^{+37}_{-35}$&$0.41^{+0.06}_{-0.06}$&$34.9^{+0.1}_{-0.3}$ &1.22&1.46(113)\\
\hline
\end{tabular}
\caption{PCA, HEXTE, SPI and ISGRI data fitted simultaneously 
using the XSPEC multicomponent model PHABS*(COMPPS+GAUSSIAN).
Gaussian line was fixed at an energy of 6.4 keV
with a width fixed to 0.1 keV. The seed photon temperature kT$_{seed}$  was frozen to 390 eV. 
W$_{Fe}$ is the equivalent width. $\Omega$/${2\pi}$ is the reflection fraction.
L$_{2-600}$  is the luminosity of the
 source in the 2-600 keV energy range. We show for the dataset of revolution 175, the fit parameters in the case for which 
the HEXTE data have been included (175*).}
\vspace*{-0.3 cm} 
\label{tab:table8bis}
\end{table} 

%% file: tab6.tex
\begin{table}[ht]
\renewcommand{\arraystretch}{0.01}
%\centering
\begin{tabular}{lccccccccc} 
\hline
rev   &  $\rm l_h/l_s$          & $\rm l_{nth}/l_{th}$     &$\tau _{es}$                  &W$_{Fe}$        &$\Omega$/${2\pi}$&$\tau _{tot}$ & kT&y&$\chi ^2$(dof)\\	
      &                         &                         &                        & eV             &                 &       & keV&           &  \\
\hline
\hline
166+167	&$5.74^{+0.19}_{-0.32}$ & 0 fr &$1.15^{+0.05}_{-0.11}$&$77^{+30}_{-40}$&$0.30^{+0.03}_{-0.04}$&1.15&98&0.88&0.94(173)\\
166+167	&$6.32^{+0.10}_{-0.10}$ & 0 fr &1.60 fr  &$83^{+50}_{-46}$&$0.26^{+0.03}_{-0.03}$&1.60&72&0.90&0.98(174)\\
174	&$5.08^{+0.17}_{-0.12}$ & 0 fr &$1.61^{+0.08}_{-0.07}$&$97^{+50}_{-35}$&$0.42^{+0.03}_{-0.03}$&1.61&64&0.81&0.82(114)\\
174&$6.35^{+0.88}_{-1.11}$&$0.40^{+0.3}_{-0.3}$&$1.98^{+0.14}_{-0.18}$&$102^{+50}_{-50}$&$0.39^{+0.04}_{-0.04}$&2.00&50&0.78&0.82(113)\\
175 	&$5.09^{+0.13}_{-0.05}$ & 0 fr &$1.66^{+0.06}_{-0.03}$&$90^{+30}_{-50}$&$0.43^{+0.03}_{-0.03}$&1.60&64&0.80&1.10(99)\\
175     &$5.61^{+2.15}_{-0.16}$&$0.28^{+0.45}_{-0.17}$&$1.54^{+0.49}_{-0.12}$&$90^{+30}_{-50}$&$0.47^{+0.05}_{-0.06}$&1.56&64&0.78&1.06(98)\\
175*	&$5.08^{+0.17}_{-0.12}$ & 0 fr &$1.61^{+0.07}_{-0.02}$&$90^{+36}_{-50}$&$0.42^{+0.03}_{-0.04}$&1.61&64&0.81&1.47(114)\\
175*     &$7.08^{+0.46}_{-0.42}$&$0.65^{+0.17}_{-0.14}$&$1.92^{+0.15}_{-0.07}$  &$94^{+92}_{-59}$&$0.45^{+0.06}_{-0.05}$&1.93&50&0.76&1.45(113)\\
\hline
\end{tabular}
\caption{PCA, HEXTE, SPI and ISGRI data fitted simultaneously using EQPAIR model combined
with a Gaussian iron line. See text for the parameters description. kT$_{seed}$ has been frozen to 390 eV, with an inclination angle of 50 degrees. 
$\tau _{tot}$ and kT are calculated from fitted parameters. We show for the dataset of revolution 175, the fit parameters in the case for which 
the HEXTE data have been included (175*).}
\vspace*{-0.3 cm} 
\label{tab:table20}
\end{table}